# A path integral methodology for obtaining thermodynamic properties of nonadiabatic systems using Gaussian mixture distributions

Neil Raymond, Dmitri Iouchtchenko, Pierre-Nicholas Roy,[a] and Marcel Nooijen[b]
*Department of Chemistry, University of Waterloo, Waterloo, Ontario N2L 3G1, Canada*

We introduce a new path integral Monte Carlo method for investigating nonadiabatic systems in thermal equilibrium and demonstrate an approach to reducing stochastic error. We derive a general path integral expression for the partition function in a product basis of continuous nuclear and discrete electronic degrees of freedom without the use of any mapping schemes. We separate our Hamiltonian into a harmonic portion and a coupling portion; the partition function can then be calculated as the product of a Monte Carlo estimator (of the coupling contribution to the partition function) and a normalization factor (that is evaluated analytically). A Gaussian mixture model is used to evaluate the Monte Carlo estimator in a computationally efficient manner. Using two model systems, we demonstrate our approach to reduce the stochastic error associated with the Monte Carlo estimator. We show that the selection of the harmonic oscillators comprising the sampling distribution directly affects the efficiency of the method. Our results demonstrate that our path integral Monte Carlo method's deviation from exact Trotter calculations is dominated by the choice of the sampling distribution. By improving the sampling distribution, we can drastically reduce the stochastic error leading to lower computational cost. *Published by AIP Publishing.* https://doi.org/10.1063/1.5025058

## I. INTRODUCTION

The Born–Oppenheimer approximation (BOA) is ubiquitous in computational quantum chemistry. In systems with well-spaced potential energy surfaces (PESs) (on the order of a few eV), the BOA is appropriate and a single PES can accurately describe the atomic motion.[1] In such cases, thermal properties for sizeable gas phase molecules are typically calculated using the harmonic oscillator (HO), rigid rotor, and ideal gas approximations.[2,3] The electronic structure calculations used to obtain the electronic energy and force constant matrix are computationally expensive, but subsequent evaluation of thermal properties is trivial. However, in systems where two or more PESs approach each other energetically, off-diagonal terms in the Hamiltonian increase in magnitude and the BOA is no longer a valid approximation.[4] These systems or regions are referred to as *nonadiabatic*. Nonadiabatic dynamics describe many important chemical reactions such as photo-induced biological processes and charge transfer in materials.[5–9]

As the aforementioned approximations break down in nonadiabatic regions, new approaches are necessary to compute properties for these systems. In the present work, we focus on developing a path integral (PI) framework to obtain the time-independent properties at thermal equilibrium, including the partition function $Z$, internal energy $U$, heat capacity $C_v$, and Gibbs energy $G$. We ultimately aim to obtain thermal properties for nonadiabatic systems with a similar accuracy to what is possible in the current single-surface methods and with modest computational effort. If we could, for example, accurately and efficiently determine the change in Gibbs energy for nonadiabatic catalytic routes, this would assist in improving the product yield and selectivity and contribute to the design of improved catalysts.[10]

Typical PI approaches used to investigate quantum mechanical systems require modification to accurately describe electronically nonadiabatic systems. Recently, there has been much interest in obtaining canonical quantum time-correlation functions for electronically nonadiabatic systems, using approaches such as initial value representation (IVR),[11–14] quantum-classical path integral (QCPI),[15] centroid molecular dynamics (CMD),[16,17] and ring polymer molecular dynamics (RPMD).[18] Numerous extensions of RPMD have been developed, such as coherent-state mapping (CS)-RPMD,[19] mean field (MF)-RPMD,[20] kinetically constrained (KC)-RPMD,[21] and mapping-variable (MV)-RPMD.[22] Many of these methods use the Meyer–Miller–Stock–Thoss (MMST) representation[13,23–26] or a variation of the MMST. These methods focus on obtaining properties such as transport properties, rates of molecular processes, and spectra. By contrast, our goal is to obtain thermal equilibrium properties for nonadiabatic systems in a computationally efficient manner, and this motivates the development of a new PI method.

Two common approaches for describing nonadiabatic systems are the adiabatic and the diabatic representations. In the adiabatic representation, the electronic wavefunctions are eigenfunctions of the electronic Hamiltonian,[27] whereas in the diabatic representation, they are approximated[28] as geometrically dependent linear combinations of specific adiabatic wavefunctions.[29–32] The diabatic representation is commonly chosen because the nuclear kinetic energy can be assumed to be diagonal,[29] while the elements of the potential energy matrix are smooth functions of nuclear geometry. In this work, we are interested in the potential energy matrix in a limited low-energy region of nuclear configurations, where a low-order Taylor series expansion can be used to describe the nonadiabatic coupling terms (NACTs).[29,33]

[a]Electronic mail: pnroy@uwaterloo.ca
[b]Electronic mail: nooijen@uwaterloo.ca

Our approach focuses solely on the investigation of electronically nonadiabatic systems that are described by a vibronic Hamiltonian which is a Hamiltonian in the diabatic representation. Vibronic models are commonly used to describe nonadiabatic effects and dynamics in spectroscopy.[34–39] Our implementation makes use of the common adiabatic-to-diabatic transformation (ADT)[33] to obtain the Hamiltonian we use in our path integral Monte Carlo (PIMC) calculations. We work directly with continuous nuclear coordinates and discrete electronic states without employing any mapping schemes such as the MMST representation. We employ an exact (in the appropriate limit) PI discretization, comparable to the state-space-based path integral (SS-PI) discretization recently proposed in Ref. [40].

A key step in our approach is to separate the full vibronic Hamiltonian into a harmonic operator $\hat{h}$ and a coupling operator $\hat{V}$. Prior studies have employed various partitionings of the Hamiltonian, primarily focusing on explicitly evaluating the kinetic energy component.[41–45] Here we choose a partitioning that will allow the full partition function to be expressed as a product of two factors: the normalization factor of a distribution $\varrho$ (that is evaluated analytically) and a Monte Carlo estimator (of the coupling contribution to the partition function). We stochastically evaluate the Monte Carlo estimator using a Gaussian mixture model (GMM), where the Gaussian mixture distribution (GMD) of the model is $\varrho$. Monte Carlo methods have previously been applied to the calculation of $Z$;[46] however, the use of a GMM, which allows for sampling without rejection, is novel. A powerful property of GMMs is the ability to form smooth approximations to arbitrarily shaped densities. GMMs maintain many of the computational and theoretical benefits of Gaussian models, making them practical for efficiently modeling higher dimensional data sets for applications such as feature extraction from speech data[47] and tracking multiple objects in digital videos.[48] Our partitioning of the Hamiltonian is therefore motivated by the computational benefits provided by GMMs.

This paper is organized as follows. Our expression for the partition function is derived in Sec. II. The method of evaluating this expression is explained in Sec. III. Section IV outlines the current implementation of this method. Results of our method are presented in Sec. V, and we discuss the conclusions in Sec. VI.

## II. PATH INTEGRAL FORMULATION

We begin by explaining the separation of our vibronic Hamiltonian

$$\hat{H} = \hat{T} + \hat{U}. \quad (1)$$

We define the harmonic operator $\hat{h}$, where

$$\hat{h} = \hat{T} + \hat{U}^{ho}, \quad (2)$$

to be diagonal in the diabatic basis with $A$ discrete electronic surfaces and $d$ nuclear co-ordinates. With this representation, we assume a vibrational kinetic energy term $\hat{T}$ that has a normal mode form. A subset of the harmonic terms on the diagonal of $\hat{U}$ in the diabatic basis form the operator $\hat{U}^{ho}$. The exact composition of $\hat{h}$ is flexible, based on the elements of the subset, allowing for optimizations to specific applications. We define the coupling operator $\hat{V}$ as the remaining components of $\hat{U}$,

$$\hat{V} = \hat{U} - \hat{U}^{ho}, \quad (3)$$

allowing us to separate $\hat{H}$ into $\hat{h}$ and $\hat{V}$,

$$\hat{H} = \hat{h} + \hat{V}. \quad (4)$$

The canonical partition function is obtained from the trace of the Boltzmann operator,

$$Z = \text{Tr}\left[e^{-\beta \hat{H}}\right], \quad (5)$$

where $\beta = (k_B T)^{-1}$ is the reciprocal temperature. We represent the nuclear configurations using the vector

$$\mathbf{R} = [R_1, R_2, \ldots, R_d] \quad (6)$$

and the electronic surfaces using $a$, where

$$a \in \{1, 2, \ldots, A\}. \quad (7)$$

The resolution of the identity for this space can be expressed as

$$\int d\mathbf{R} \sum_{a=1}^{A} |\mathbf{R}, a\rangle \langle \mathbf{R}, a|. \quad (8)$$

By construction, $\hat{h}$ is diagonal in the electronic surfaces $a$,

$$\langle \mathbf{R}, a|\hat{h}|\mathbf{R}', a'\rangle = \langle \mathbf{R}|\hat{h}^a|\mathbf{R}'\rangle \delta_{aa'}, \quad (9)$$

and $\hat{V}$ is diagonal in the nuclear configurations $\mathbf{R}$,

$$\langle \mathbf{R}, a|\hat{V}|\mathbf{R}', a'\rangle = \langle a|\hat{V}(\mathbf{R})|a'\rangle \delta(\mathbf{R} - \mathbf{R}'). \quad (10)$$

Applying the symmetric Trotter factorization[49] where $P$ is the number of imaginary time-slices, also known as "beads," and $\tau = \frac{\beta}{P}$ results in

$$e^{-\beta \hat{H}} = \lim_{P \to \infty} \left(e^{-\tau \hat{h}} e^{-\tau \hat{V}}\right)^P. \quad (11)$$

Repeated insertion of the resolution of the identity yields a PI discretization of the partition function

$$Z = \lim_{P \to \infty} \int d\mathbf{R}^P \sum_{\mathbf{a}} \prod_{i=1}^{P} \langle \mathbf{R}_i, a_i|e^{-\tau \hat{h}}|\mathbf{R}_{i+1}, a_i\rangle$$
$$\times \langle \mathbf{R}_{i+1}, a_i|e^{-\tau \hat{V}}|\mathbf{R}_{i+1}, a_{i+1}\rangle. \quad (12)$$

We make use of the compact notations

$$\sum_{\mathbf{a}} = \sum_{a_1=1}^{A} \sum_{a_2=1}^{A} \cdots \sum_{a_P=1}^{A} \quad (13)$$

and

$$\int d\mathbf{R}^P = \int d\mathbf{R}_1 \int d\mathbf{R}_2 \cdots \int d\mathbf{R}_P. \quad (14)$$

The general expression for Z in Eq. (12) allows us to evaluate $e^{-\tau \hat{h}}$ independently of $e^{-\tau \hat{V}}$. We will show later that $e^{-\tau \hat{h}}$ can be evaluated analytically.

A finite choice of P results in a systematic error in the Trotter factorization. For readability, we suppress this approximation for the remainder of this paper. As our calculations use finite values for P, this error is present in all of the work that follows, but we systematically reduce it by increasing P.

Additionally we define

$$g(\mathbf{R}) = \sum_{\mathbf{a}}^{A} \prod_{i=1}^{P} \langle \mathbf{R}_i, a_i | e^{-\tau \hat{h}} | \mathbf{R}_{i+1}, a_i \rangle \langle \mathbf{R}_{i+1}, a_i | e^{-\tau \hat{V}} | \mathbf{R}_{i+1}, a_{i+1} \rangle \quad (15)$$

which can be treated as a probability density function (PDF)[50] with the normalization factor

$$Z = \int d\mathbf{R}^P g(\mathbf{R}). \quad (16)$$

For brevity, we use the convention that $\varrho(\mathbf{R}) = \varrho(\mathbf{R}_1, \mathbf{R}_2, \ldots, \mathbf{R}_P)$ and $g(\mathbf{R}) = g(\mathbf{R}_1, \mathbf{R}_2, \ldots, \mathbf{R}_P)$. In this usage, **R** is purely a shorthand notation and should not be confused with the **R** of Eq. (6).

## III. OBTAINING THE PARTITION FUNCTION

We have demonstrated how to derive a general PI expression for Z in nuclear coordinates **R** and electronic surfaces $a$. Next, we show how we evaluate Z as a product of a Monte Carlo estimator and a normalization factor. We apply two stochastic methods: Monte Carlo integration, to avoid directly evaluating the integrals over **R** due to the computational cost; and importance sampling, to reduce the variance by sampling from a new distribution $\varrho$. We use the following convention: for a PDF $f(x)$, we treat the corresponding symbol $f$ as the distribution defined by $f(x)$.

Sampling from $\varrho$ with weight $\frac{g}{\varrho}$ is equivalent to sampling from the original distribution $g$, biasing the sample obtained toward $\varrho$.[51] If $\varrho$ represents the dominant contributions to $g$, this results in a reduction of variance, which leads to more efficient calculation of parameters of interest, since fewer samples are needed for convergence. For an arbitrary PDF $\varrho(\mathbf{R})$, from

$$\frac{Z}{\int d\mathbf{R}^P \varrho(\mathbf{R})} = \frac{\int d\mathbf{R}^P \varrho(\mathbf{R}) \frac{g(\mathbf{R})}{\varrho(\mathbf{R})}}{\int d\mathbf{R}^P \varrho(\mathbf{R})}, \quad (17)$$

it follows that

$$Z = \left\langle \frac{g(\mathbf{R})}{\varrho(\mathbf{R})} \right\rangle_\varrho \left( \int d\mathbf{R}^P \varrho(\mathbf{R}) \right). \quad (18)$$

We define the normalization of the PDF $\varrho(\mathbf{R})$ to be

$$Z_\varrho = \int d\mathbf{R}^P \varrho(\mathbf{R}) \quad (19)$$

which will take on the role of a partition function in the following due to our choice of $\varrho$. We define $Z_{MC}$ as the estimate of the coupling contribution to Z,

$$Z_{MC} = \left\langle \frac{g(\mathbf{R})}{\varrho(\mathbf{R})} \right\rangle_\varrho, \quad (20)$$

resulting in a compact representation for the partition function of our Hamiltonian

$$Z = Z_{MC} Z_\varrho. \quad (21)$$

We choose a distribution $\varrho$ that can be sampled without rejection and whose normalization can be analytically evaluated. The partition function is therefore the product of an estimate $Z_{MC}$, obtained using Monte Carlo integration, and a normalization factor $Z_\varrho$.

### A. Matrix representations of propagators

We introduce the following notation for the matrix representation of the harmonic and coupling propagators. We define the matrices $\mathbb{O}$ and $\mathbb{M}$ through their matrix elements,

$$\mathbb{O}(\mathbf{R}, \mathbf{R}')_{aa'} = \langle \mathbf{R} | e^{-\tau \hat{h}^a} | \mathbf{R}' \rangle \delta_{aa'} \quad (22)$$

and

$$\mathbb{M}(\mathbf{R})_{aa'} = \langle a | e^{-\tau \hat{V}(\mathbf{R})} | a' \rangle. \quad (23)$$

This allows us to express Eq. (15) in terms of the matrices $\mathbb{O}$ and $\mathbb{M}$,

$$g(\mathbf{R}) = \mathrm{Tr}\left[ \prod_{i=1}^{P} \mathbb{O}(\mathbf{R}_i, \mathbf{R}_{i+1}) \mathbb{M}(\mathbf{R}_{i+1}) \right], \quad (24)$$

where the trace is over the electronic degrees of freedom (DoFs). The matrix $\mathbb{M}$ is evaluated through diagonalization of the matrix $V(\mathbf{R})_{aa'}$ at a given configuration **R**. The matrix $\mathbb{O}$ is evaluated analytically which will be shown in Sec. IV.

The PDF $g(\mathbf{R})$ describes a system with intersurface coupling. For most systems, this PDF is computationally difficult to evaluate and infeasible to sample directly. Therefore, we consider a PDF $\varrho(\mathbf{R})$ that omits the intersurface coupling,

$$\varrho(\mathbf{R}) = \mathrm{Tr}\left[ \prod_{i=1}^{P} \mathbb{O}(\mathbf{R}_i, \mathbf{R}_{i+1}) \right]. \quad (25)$$

$\varrho(\mathbf{R})$ is defined by the diagonal matrix $\mathbb{O}$ of order $A$ and therefore by the harmonic portion $\hat{h}$ of the system's Hamiltonian. The corresponding distribution $\varrho$ is a sum of multi-dimensional Gaussian distributions and thus has the important property of being a Gaussian mixture distribution (GMD).[52] This property of $\varrho(\mathbf{R})$ is key to our PIMC method. A GMD is less complex than $g(\mathbf{R})$ and is computationally efficient to evaluate. In Sec. IV C, we will consider a more general PDF whose corresponding distribution is also a GMD but whose parameters are independent of the system's Hamiltonian which allows for greater statistical accuracy.

## IV. IMPLEMENTATION

Our current algorithm operates on a vibronic Hamiltonian in normal mode co-ordinates,

$$\hat{H}^{aa'} = E^{aa'} + \left( \frac{1}{2} \sum_{j}^{N} \omega_j \left( \hat{p}_j^2 + \hat{q}_j^2 \right) \right) \delta_{aa'} + \sum_{j}^{N} g_j^{aa'} \hat{q}_j + \frac{1}{2} \sum_{jj'}^{N} G_{jj'}^{aa'} \hat{q}_j \hat{q}_{j'}, \quad (26)$$

where the dimensionless nuclear configuration representations are labeled $q$ instead of **R** and all parameters have units of energy ($\hbar = 1$). Recall that the composition of $\hat{h}$ was dependent on the choice of $\hat{U}^{ho}$. In our implementation, we have selected $\hat{U}^{ho}$ such that the surface-dependent harmonic operator is defined to be

$$\hat{h}^a = E^{aa} + \frac{1}{2}\sum_{j}^{N}\omega_j\left(\hat{p}_j^2 + \hat{q}_j^2\right) + \sum_{j}^{N} g_j^{aa}\hat{q}_j. \quad (27)$$

This is analogous to the shifted bath Hamiltonian in the quasi-adiabatic propagator path integral (QuAPI) method,[43–45] described by Eq. (9c) in Ref. 43. The harmonic operator has the same frequencies for each surface but can differ in energy and in displacement of the normal modes. Within our framework, this restriction is made for computational efficiency and can be lifted in principle. However, in Ref. 43, this restriction is inherent to the system-bath Hamiltonian. For systems where the quadratic terms on the diagonal of $\hat{H}$ are significant, the operator $\hat{h}^a$ may also be extended to include them, but this incurs a computational penalty due to the need for a different co-ordinate rotation for each surface.

### A. Analytical representation of $\varrho$

We begin by deriving the analytical form of the diagonal matrix $\mathbb{O}$ and the PDF $\varrho(\boldsymbol{q})$. Consider the propagator

$$\langle x|e^{-\tau\hat{h}_o}|x'\rangle, \quad (28)$$

where $\hat{h}_o$ has the form of a one-dimensional quantum harmonic oscillator (QHO) in natural length co-ordinates and frequency $\omega_j$. The analytical expression for this propagator is[53]

$$K(x, x'; \tau\omega_j) = \mathcal{F}_j \exp\left(\mathcal{S}_j x x' - \mathcal{C}_j \frac{1}{2}\left(x^2 + (x')^2\right)\right), \quad (29)$$

where

$$\mathcal{C}_j = \coth(\tau\omega_j), \quad (30)$$
$$\mathcal{S}_j = \operatorname{csch}(\tau\omega_j), \quad (31)$$
$$\mathcal{F}_j = \sqrt{\frac{\mathcal{S}_j}{2\pi}}. \quad (32)$$

The surface-dependent harmonic operator for multiple normal modes is expressed by completing the square as

$$\hat{h}^a = \left[E^{aa} + \Delta^a\right] + \left[\frac{1}{2}\sum_{j=1}^{N}\omega_j\left(\hat{p}_j^2 + (\hat{x}_j^a)^2\right)\right] \quad (33a)$$
$$= \tilde{E}^a + \hat{h}_o^a, \quad (33b)$$

where

$$x_j^a = q_j - d_j^a, \quad (34)$$
$$d_j^a = \frac{-g_j^{aa}}{\omega_j}, \quad (35)$$
$$\Delta^a = -\frac{1}{2}\sum_{j=1}^{N}\frac{(g_j^{aa})^2}{\omega_j}. \quad (36)$$

Expressing the matrix elements of $\mathbb{O}$ in terms of $K$ as

$$\mathbb{O}(\boldsymbol{q}_i, \boldsymbol{q}_{i+1})_{aa} = \langle q_{1,i} q_{2,i} \cdots q_{N,i}|e^{-\tau\hat{h}^a}|q_{1,i+1} q_{2,i+1} \cdots q_{N,i+1}\rangle \quad (37a)$$
$$= \left(e^{-\tau\tilde{E}^a}\right)\langle x_{1,i}^a x_{2,i}^a \cdots x_{N,i}^a|e^{-\tau\hat{h}_o^a}|x_{1,i+1}^a x_{2,i+1}^a \cdots x_{N,i+1}^a\rangle \quad (37b)$$
$$= \left(e^{-\tau\tilde{E}^a}\right)\prod_{j=1}^{N} K\left(x_{j,i}^a, x_{j,i+1}^a; \tau\omega_j\right), \quad (37c)$$

we can analytically evaluate $\varrho(\boldsymbol{q})$ as follows:

$$\varrho(\boldsymbol{q}) = \sum_{a=1}^{A}\left(e^{-\beta\tilde{E}^a}\right)\prod_{j=1}^{N}\left(\mathcal{F}_j\right)^P \pi_j(\boldsymbol{x}_j^a), \quad (38)$$

where

$$\pi_j(\boldsymbol{x}_j^a) = \exp\left[\mathcal{S}_j\sum_{i=1}^{P} x_{j,i}^a x_{j,i+1}^a - \mathcal{C}_j\sum_{i=1}^{P}\left(x_{j,i}^a\right)^2\right] \quad (39)$$

and

$$\boldsymbol{x}_j^a = \left[x_{j,1}^a, x_{j,2}^a, \ldots, x_{j,P}^a\right]. \quad (40)$$

Note that we employ the same convention for $\varrho(\boldsymbol{q})$ and $g(\boldsymbol{q})$ as we do for $\varrho(\boldsymbol{R})$ and $g(\boldsymbol{R})$ that $\varrho(\boldsymbol{q}) = \varrho(\boldsymbol{q}_1, \boldsymbol{q}_2, \ldots, \boldsymbol{q}_P)$ and $g(\boldsymbol{q}) = g(\boldsymbol{q}_1, \boldsymbol{q}_2, \ldots, \boldsymbol{q}_P)$.

### B. Derivation of sampling distribution $\varrho$

The general form of the PDF of a mixture distribution is a convex combination of PDFs $p_i(x)$,

$$f(x) = \sum_{i} w_i p_i(x), \quad (41)$$

where $\sum_i w_i = 1$ and $w_i \geq 0$ for all $i$. Using the analytical expression of $\varrho(\boldsymbol{q})$, we will show that it is a PDF of a GMD which is a mixture distribution where each $p_i(x)$ represents a Gaussian distribution.

We start by re-expressing $\pi$ in quadratic form,

$$\pi_j(\boldsymbol{x}_j^a) = \exp\left[-\frac{1}{2}(\boldsymbol{x}_j^a)^\top\left(2\mathcal{C}_j\underline{\mathbb{1}} - \mathcal{S}_j\underline{\boldsymbol{B}}\right)\boldsymbol{x}_j^a\right], \quad (42)$$

where $\underline{\boldsymbol{B}}$ is a circulant matrix of dimension $P \times P$ defined by the row vector

$$[0\ 1\ 0\ 0\ \cdots\ 0\ 0\ 1]. \quad (43)$$

Then, we define the PDFs of $A$ multivariate Gaussians

$$\varrho^a(\boldsymbol{q}) = \frac{1}{Z_{\varrho^a}} \prod_{j=1}^{N} (\mathcal{F}_j)^P \pi_j(\boldsymbol{x}_j^a), \tag{44}$$

where

$$Z_{\varrho^a} = \prod_{j=1}^{N} \frac{1}{2} \text{csch}\left(\frac{\beta \omega_j}{2}\right). \tag{45}$$

We express $\varrho(\boldsymbol{q})$ as

$$\varrho(\boldsymbol{q}) = \sum_{a=1}^{A} w^a \varrho^a(\boldsymbol{q}), \tag{46}$$

with weights

$$w^a = \frac{e^{-\beta \tilde{E}^a} Z_{\varrho^a}}{Z_\varrho}, \tag{47}$$

where

$$Z_\varrho = \sum_{a=1}^{A} e^{-\beta \tilde{E}^a} Z_{\varrho^a}. \tag{48}$$

We can see from Eqs. (41) and (46) that $\varrho$ satisfies the definition of a GMD with means $\boldsymbol{d}^a$ and covariance matrices

$$\underline{\boldsymbol{\Sigma}}_j = \left(2\mathcal{C}_j \underline{\mathbb{1}} - \mathcal{S}_j \underline{\boldsymbol{B}}\right)^{-1}. \tag{49}$$

To sample from this GMD efficiently, we must decouple the bead DoFs. To do this, we diagonalize $\underline{\boldsymbol{B}}$ by the unitary transformation $\underline{\boldsymbol{\mathcal{V}}}$ so that

$$B_{ii'} = \sum_{\lambda=1}^{P} \mathcal{V}_{i\lambda} b_{\lambda\lambda} \mathcal{V}_{i'\lambda}^*. \tag{50}$$

This allows us to define collective bead co-ordinates,

$$y_{j\lambda}^a = \sum_{i=1}^{P} x_{ji}^a \mathcal{V}_{i\lambda}, \tag{51}$$

which are uncoupled, leading to straightforward sampling from one-dimensional Gaussians.

### C. Generalization of $\varrho$

We will now refer to the previously discussed sampling distribution as $\varrho_0$ and use $\varrho$ to indicate a generalized distribution. We have defined the distribution $\varrho_0$ in terms of the diagonal matrix $\mathbb{O}$ of order $A$ and therefore the harmonic portion $\hat{h}$ of the system's Hamiltonian. The distribution $\varrho_0$ is fully specified by the system's parameters $\omega_j$, $E^{aa}$, and $g_j^{aa}$. This definition can be generalized to a distribution

$$\varrho(\boldsymbol{q}) = \text{Tr}\left[\prod_{i=1}^{P} \tilde{\mathbb{O}}(\boldsymbol{q}_i, \boldsymbol{q}_{i+1})\right], \tag{52}$$

defined in terms of a diagonal matrix $\tilde{\mathbb{O}}$ of order $\tilde{A}$ expressed in terms of an operator $\hat{h}$ of the form in Eq. (27), whose parameters are independent of the system's Hamiltonian. In Sec. V, we use a subscript index to differentiate between different distributions: $\varrho_1$ and $\varrho_2$, for example. We reserve the index 0 to refer to the distribution defined by the harmonic portion of the system's Hamiltonian. Of particular importance is the distinction that the number of multivariate Gaussians $\tilde{A}$ comprising $\varrho$ may differ from the number of electronic states $A$ present in the system's Hamiltonian. In fact, the value of $\tilde{A}$ may be chosen arbitrarily for each calculation, and appropriate values will in general be system-dependent.

The first implication of this generalization arises in the evaluation of $Z_{\text{MC}}$. If one chooses a distribution $\varrho \neq \varrho_0$ as their sampling distribution, then the ratio in Eq. (20) contains $\tilde{\mathbb{O}}$ matrices in the denominator rather than $\mathbb{O}$ matrices. This roughly doubles the computational cost of evaluating these matrices over sampled points compared to the case with $\varrho_0$, where the same matrices are present in both the numerator and denominator. Note that the evaluation of the $\mathbb{O}$ matrices constitutes roughly 10%–20% of the total computational cost in the present implementation, and as evident in Sec. V, the improved statistical accuracy provided by different choices of $\varrho$ vastly outweighs the additional computation required. The second implication is that the selection of $\varrho$ reduces to a distribution fitting or parameter estimation problem, which is commonly addressed using statistical methods such as maximum likelihood estimation (MLE).[54]

### D. Evaluation algorithm

The general algorithm used to calculate $Z$ is presented. Consider a system described by a vibronic Hamiltonian of the form in Eq. (26) which has $N$ normal modes and $A$ electronic states. This system is evaluated at a specific temperature $\beta$, the number of Monte Carlo samples $L$, the number of beads $P$, and the sampling distribution $\varrho$. We will explain the sampling and evaluation processes for the $L = 2$ case. Because all samples are independent, the generalization to $L > 2$ is trivial.

First, we draw two random values $\tilde{a}_1$, $\tilde{a}_2$ from a discrete distribution from 1 to $\tilde{A}$ with weights $[w^1, w^2, \ldots, w^{\tilde{A}}]$. These variables $\tilde{a}_\ell$ determine from which of the $\tilde{A}$ multivariate Gaussians $\varrho^{\tilde{a}}$ (that comprise the sampling distribution $\varrho$) each individual sample is drawn. Next, we generate samples $\boldsymbol{y}^1, \boldsymbol{y}^2$ of dimension $N \times P$ from the normal distributions $\mathcal{N}_{N,P}(\boldsymbol{d}^{\tilde{a}_1}, \underline{\boldsymbol{\Sigma}})$, and $\mathcal{N}_{N,P}(\boldsymbol{d}^{\tilde{a}_2}, \underline{\boldsymbol{\Sigma}})$. We transform from collective bead co-ordinates $y_{j\lambda}^\ell$ to bead dependent co-ordinates $x_{ji}^\ell$,

$$\underline{\boldsymbol{x}}^\ell = \underline{\boldsymbol{y}}^\ell (\underline{\boldsymbol{\mathcal{V}}})^\dagger, \tag{53}$$

where $\underline{\boldsymbol{\mathcal{V}}}$ is defined in Eq. (50). To evaluate the matrices $\mathbb{O}$ and $\mathbb{M}$, we shift each sample to all $A$ electronic states,

$$x_{ji}^{\ell,a} = x_{ji}^\ell + d_j^{\tilde{a}_\ell} - d_j^a, \tag{54}$$

resulting in a tensor $\boldsymbol{x}$ of dimension $2 \times A \times N \times P$. To evaluate the matrices $\tilde{\mathbb{O}}$, we shift each sample to all $\tilde{A}$ fictitious states,

$$\tilde{x}_{ji}^{\ell,\tilde{a}} = x_{ji}^\ell + d_j^{\tilde{a}_\ell} - d_j^{\tilde{a}}, \tag{55}$$

resulting in a tensor $\tilde{\boldsymbol{x}}$ of dimension $2 \times \tilde{A} \times N \times P$. In this manner, we evaluate $\varrho(\boldsymbol{q})$ and $g(\boldsymbol{q})$ for each sample using Eqs. (24) and (52). Finally, $Z$ [Eq. (21)] is the product of $Z_{\text{MC}}$ [Eq. (20)],

$$Z_{\text{MC}} = \frac{1}{L} \sum_{\ell=1}^{L} \frac{g(\boldsymbol{q}^\ell)}{\varrho(\boldsymbol{q}^\ell)}, \tag{56}$$

and $Z_\varrho$ [Eqs. (45) and (48)],

$$Z_\varrho = \sum_{\tilde{a}=1}^{\tilde{A}} \left(e^{-\beta \tilde{E}^{\tilde{a}}}\right) \prod_{j=1}^{N} \frac{1}{2}\text{csch}\left(\frac{\beta\omega_j}{2}\right). \quad (57)$$

Note that in our current implementation we draw and evaluate samples $y$ in blocks of size $L' < L$ which is dependent on the dimensionality of the system. This is to exploit specific hardware architecture (spatial and temporal locality) of general purpose central processing units (CPUs) for increased computational performance.[55–57] Preliminary testing has shown that this block-based sampling approach can provide a computational runtime improvement of two orders of magnitude.

### E. Systematic Trotter error for vibronic Hamiltonian

The following is our approach for calculating the partition function with the inclusion of systematic Trotter error. The approach is in the spirit of the iterative scheme presented in Ref. 58 but is for a Hamiltonian in the diabatic representation that contains multiple discrete electronic states.

The Trotter-factorized partition function is

$$Z_{\text{Trotter}}(P) = \text{Tr}\left(e^{-\tau\hat{h}}e^{-\tau\hat{V}}\right)^P. \quad (58)$$

We may express the operators $\hat{h}$ and $\hat{V}$ in matrix form in a HO basis set augmented with an electronic state label. This basis has the states

$$|v\rangle = |n_1, n_2, \ldots, n_N, a\rangle, \quad (59)$$

where $n_i$ denotes the number of quanta in mode $i$ and $a$ labels the electronic state. Because in practice the basis will always be truncated with respect to the number of quanta of vibration, the resulting matrices $\underline{h}$ and $\underline{V}$ are technically approximate representations of the respective operators. However, we take care to ensure that the basis is large enough for satisfactory convergence.

We construct the matrix

$$\underline{M} = e^{-\tau\underline{h}}e^{-\tau\underline{V}} \quad (60)$$

by first taking the matrix exponential of each matrix. It is then possible to efficiently obtain

$$Z_{\text{Trotter}}(P) = \text{Tr}\underline{M}^P \quad (61)$$

without performing any matrix multiplications by diagonalizing $\underline{M}$ to obtain its eigenvalues $\alpha_u$ and then computing

$$Z_{\text{Trotter}}(P) = \sum_u (\alpha_u)^P. \quad (62)$$

Having access to this value allows us to gauge the correctness and accuracy of our PIMC implementation.

Since we have the matrices $\underline{h}$ and $\underline{V}$, we may also generate the Hamiltonian matrix

$$\underline{H} = \underline{h} + \underline{V}. \quad (63)$$

Direct diagonalization of this matrix produces the exact energy eigenvalues $E_n$ of the full coupled system which we may use in the sum-over-states (SOS)

$$Z_{\text{SOS}} = \sum_n e^{-\beta E_n} \quad (64)$$

to calculate the exact value of the partition function. This quantity does not include any Trotter error and is used as the reference value for the $\tau \to 0$ limit in the model system results.

### F. Computational libraries

Two computational libraries, Pibronic[59] and VibronicToolkit,[60] were developed alongside the research that is presented in this paper. Pibronic contains the full implementation of our method. VibronicToolkit is a proof of concept implementation of our method and was used to verify the results from Pibronic. Both of these libraries are open-source and available on GitHub.[59,60]

## V. RESULTS

To investigate the effectiveness of our PIMC method, the results for two systems are presented: "Displaced" and "Jahn–Teller." The Displaced system highlights the effect of $\varrho$ on the accuracy and efficiency of sampling when there are multimodal displacements. The Jahn–Teller system is representative of magnetic systems and systems containing radicals.[61,62] These systems are described by vibronic Hamiltonians with the same form as Eq. (26), but without any quadratic coupling terms ($G_{jj'}^{aa'} = 0$). To allow for numerical analysis such as SOS, the systems were restricted to two normal modes and two electronic surfaces. Both systems have a single tunable parameter and were evaluated over a range of six values. All graphics are labeled with the associated tunable parameter in the upper right-hand corner and the associated choice of $\varrho$ in the top left-hand corner. All PIMC results were calculated with one million samples ($L = 10^6$) and at 300 K. The two properties of interest are $\tau$ convergence of our PIMC results and the suitability of $\varrho$ as a sampling distribution for $g$.

For both systems, we present the Hamiltonian, the system's parameters, and two plots: an elevation map of the lower PES and a $\tau$ convergence plot of $Z$. In the elevation maps (Figs. 1 and 3), the minima of the diabatic surfaces comprising $\varrho$ are represented by crosses (+).

We wish to obtain a visual representation of the distribution $\varrho$ in normal mode co-ordinates $q$ so that it can be displayed in Figs. 1 and 3. Using such a representation, we can reason about the suitability of $\varrho$ as a model of $g$. Since $\varrho$ is composed of Gaussian distributions, it would be natural to use their standard deviations for this representation. However, within our formalism, the notion of variance for independent DoFs only exists in collective bead co-ordinates which cannot be readily visualized.

The distribution of the mean of the path

$$\bar{x}_j^a = \frac{1}{P}\sum_{i=1}^{P} x_{ji}^a \quad (65)$$

provides a measure of the spread of the samples in normal mode co-ordinates. Conveniently, the centroid collective bead co-ordinate

$$y_{j1}^a = \sum_{i=1}^{P} x_{ji}^a \mathcal{V}_{i1} \quad (66)$$

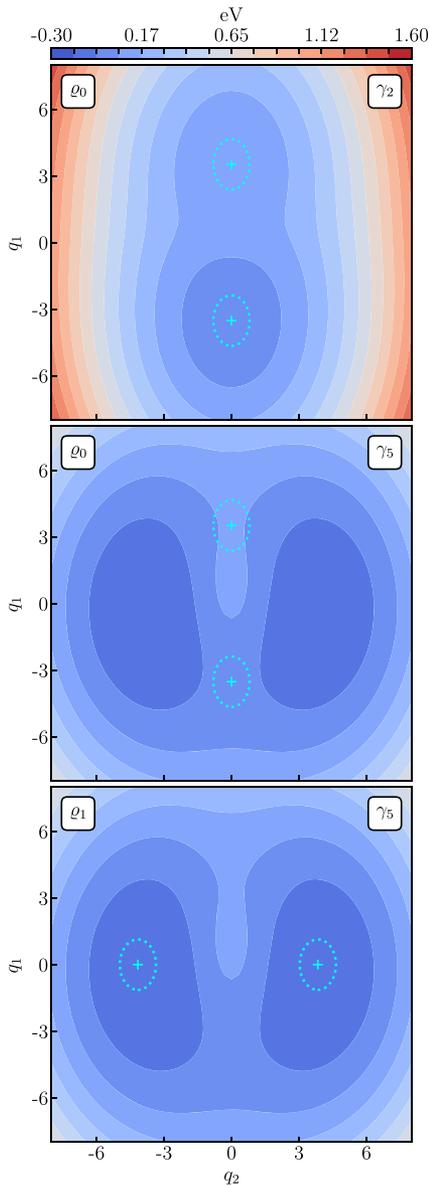

FIG. 1. Three elevation maps of the lower adiabatic PES of the Displaced system at $\gamma_2 = 0.04$ eV (top panel) and $\gamma_5 = 0.16$ eV (middle and bottom panels). The colormap for the elevation is presented at the top of the figure. For each panel, the sampling distribution $\varrho_i$ is listed in the upper left-hand corner, and an ellipse is centered on the minimum, indicated by the cross (+), of each HO comprising $\varrho_i$ as described in Sec. V.

is directly related to $\bar{x}_j^a$ because the entries of the eigenvector $\mathcal{V}_{i1}$ are all equal,

$$y_{j1}^a = \frac{1}{\sqrt{P}} \sum_{i=1}^{P} x_{ji}^a. \tag{67}$$

We know from Eqs. (49) and (50) that the standard deviation of $y_{j1}^a$ is given by

$$\sigma_j = \left(2\mathcal{C}_j - \mathcal{S}_j b_{11}\right)^{-\frac{1}{2}}, \tag{68}$$

where $b_{11}$ is the largest eigenvalue of $\underline{\mathbf{B}}$, and therefore, the standard deviation of $\bar{x}_j^a$ is

$$\tilde{\sigma}_j = \frac{\sigma_j}{\sqrt{P}}. \tag{69}$$

Note that $\tilde{\sigma}_j$ does not appear to be sensitive to changes in the number of beads $P$. We therefore choose to visually represent the distribution $\varrho$ with an ellipse centered at the minimum of each HO in Figs. 1 and 3. The diameters of each ellipse are $2\tilde{\sigma}_1$ in the $q_1$ mode and $2\tilde{\sigma}_2$ in the $q_2$ mode.

In the $\tau$ convergence plots (Figs. 2 and 4), the $y$ axis is defined as

$$\Delta Z = \left(\frac{Z - Z_{\text{SOS}}}{Z_{\text{SOS}}}\right)(100\%), \tag{70}$$

where we calculate $Z_{\text{SOS}}$ by SOS with eighty HO basis functions for each normal mode. The sum-over-states calculation is shown in Sec. IV E. It is important to realize that the sign in these plots is only representative of which parameter ($Z$ or $Z_{\text{SOS}}$) is larger.

Our PIMC method has three sources of error, not including the inevitable floating-point error associated with carrying out real-number calculations on a computer:

(i) the choice of a finite $P$ introduces systematic error due to the Trotter factorization, Eq. (12);
(ii) drawing a finite number $L$ of samples $\underline{\mathbf{y}}$ from $\varrho$ with which we evaluate our estimators; and
(iii) the choice of a sampling distribution $\varrho$ that is different from the true distribution $g$.

The most accurate estimate of a property that we can calculate for a fixed choice of $P$ includes error (i) from the Trotter factorization. This is a formal error that is present in all PI methods and cannot be eliminated. Consequently, our goal is to reduce the difference between our PIMC results and the Trotter results. Additionally, in practice, all Monte Carlo methods are restricted to a finite number of samples leading to error (ii) and a non-zero variance. Therefore, we attempt to reduce error (iii) which is introduced by our use of importance sampling. We differentiate between these sources of error by comparing our PIMC results to SOS calculations that include the Trotter error, represented by black curves in the $\tau$ convergence plots.

### A. Displaced system

This system is described by the Hamiltonian

$$\hat{H} = \hat{h} + \hat{V} \tag{71a}$$

$$= \begin{bmatrix} E^a + \hat{h}_o + \lambda \hat{q}_1 & 0 \\ 0 & E^b + \hat{h}_o - \lambda \hat{q}_1 \end{bmatrix} + \gamma_i \begin{bmatrix} 0 & \hat{q}_2 \\ \hat{q}_2 & 0 \end{bmatrix}, \tag{71b}$$

with the parameters given in Table I.

Results are analyzed as a function of $\gamma$, the strength of the coupling. At $\gamma_1$, the two PESs are displaced along the $q_1$ axis. Increasing $\gamma$ introduces a displacement along the $q_2$ axis as seen in Fig. 1.

We begin with the simple distribution $\varrho_0$, derived from $\hat{h}$. The top panel of Fig. 1 demonstrates that the HOs comprising $\varrho_0$ are a very good model of the system for a low value ($\gamma_2$) of the coupling. However, the middle panel of Fig. 1 shows that $\varrho_0$ is less reasonable for a high value ($\gamma_5$). As we increase the coupling $\gamma$, the contribution from the $q_2$ mode increases and we push the system farther into a nonadiabatic regime. A better choice of the distribution $\varrho$ is necessary.

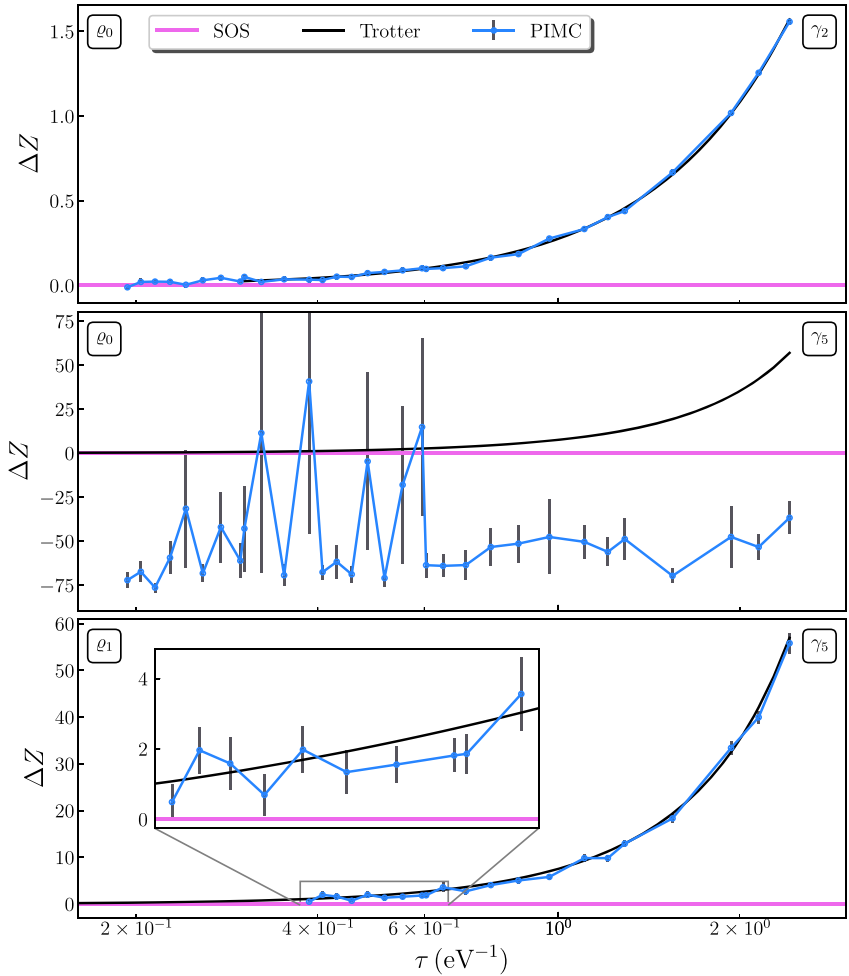

FIG. 2. Three $\tau$ convergence plots of the $\Delta Z$ deviation of our PIMC method (blue) and the exact Trotter (black) from $Z_{\text{SOS}}$ (magenta). The upper panel demonstrates that the simple distribution $\varrho_0$ is a reasonable description of the Displaced system when it is mostly harmonic ($\gamma_2$). When the system is in a nonadiabatic regime ($\gamma_5$), $\varrho_0$ is no longer appropriate as seen in the middle panel. Comparing the middle panel to the bottom panel, it is clear that changing the sampling distribution from $\varrho_0$ to $\varrho_1$ has a significantly positive impact on our results.

The obvious modification to $\varrho$ is to include a $q_2$ term. If the system is in a nonadiabatic regime, then the $q_2$ mode should be the most important contribution. However, because of the mixing of the modes, just adding a $q_2$ term is suboptimal. By looking at Fig. 1, we can see that a $q_1$ displacement similar to that in $\hat{h}$ would not be appropriate in the $\gamma_5$ regime. In addition, to reduce the variance we want to bias $\varrho$ toward the minima along the $q_2$ mode. We therefore choose a new distribution $\varrho_1$, replacing the $q_1$ term in $\hat{h}$ with a $q_2$ term,

$$\begin{bmatrix} E^a + \hat{h}_o + \gamma_i \hat{q}_2 & 0 \\ 0 & E^b + \hat{h}_o - \gamma_i \hat{q}_2 \end{bmatrix}. \quad (72)$$

Due to the importance of the $q_2$ mode, the reduced accuracy in the description of the $q_1$ mode should not have a noticeable effect on our results. As this is a direct application of importance sampling, we expect that $\tau$ convergence of $Z$ will be improved by using sampling distribution $\varrho_1$ instead of $\varrho_0$.

The results using $\varrho_1$ show a large reduction of the stochastic error compared to the results using $\varrho_0$, shown in Fig. 2. With the improved distribution $\varrho_1$, the deviation has been reduced from nearly 100% to 1%. This shows that the choice of the sampling distribution dominates our PIMC method's deviation from the exact Trotter. This choice of $\varrho_1$ is analogous to identifying the local minima of the ground state PES.

### B. Jahn–Teller system

This system is described by the Hamiltonian

$$\hat{H} = \hat{h} + \hat{V} \quad (73a)$$

$$= \begin{bmatrix} E_i + \hat{h}_o + \lambda_i \hat{q}_1 & 0 \\ 0 & E_i + \hat{h}_o - \lambda_i \hat{q}_1 \end{bmatrix} + \lambda_i \begin{bmatrix} 0 & \hat{q}_2 \\ \hat{q}_2 & 0 \end{bmatrix}, \quad (73b)$$

with the parameters given in Table II.

TABLE I. Displaced system parameters (eV).

| Parameter | |
|---|---|
| $E^a$ | 0.0996 |
| $E^b$ | 0.1996 |
| $\omega_1$ | 0.02 |
| $\omega_2$ | 0.04 |
| $\lambda$ | 0.072 |
| $\gamma_1$ | 0.00 |
| $\gamma_2$ | 0.04 |
| $\gamma_3$ | 0.08 |
| $\gamma_4$ | 0.12 |
| $\gamma_5$ | 0.16 |
| $\gamma_6$ | 0.20 |

TABLE II. Jahn–Teller system parameters (eV).

| Parameter | |
|---|---|
| $E_1$ | −0.029 99 |
| $E_2$ | −0.003 33 |
| $E_3$ | 0.076 66 |
| $E_4$ | 0.209 99 |
| $E_5$ | 0.396 67 |
| $E_6$ | 0.631 35 |
| $\omega_1$ | 0.03 |
| $\omega_2$ | 0.03 |
| $\lambda_1$ | 0.00 |
| $\lambda_2$ | 0.04 |
| $\lambda_3$ | 0.08 |
| $\lambda_4$ | 0.12 |
| $\lambda_5$ | 0.16 |
| $\lambda_6$ | 0.20 |

Results are analyzed as a function of $\lambda$, the strength of the linear terms. For $\lambda_i$ where $i > 1$, this system's ground state PES has the form of a champagne bottle. As we increase $\lambda$, the curvature of the well increases. For $\lambda_2$ in the top panel of Fig. 3, there appears to be no well because it is too small to be resolved at that scale. The energies $E_i$ were chosen so that the ground state energy $\approx 0$ eV.

The top panel of Fig. 3 demonstrates that the HOs comprising $\varrho_0$ are again a very good model of the system for a low value ($\lambda_2$) of the linear term. As with the previous system, the second panel of Fig. 3 shows that $\varrho_0$ is less reasonable for a high value ($\lambda_5$). Contrary to the Displaced system, relocating the HOs will not reduce the stochastic error due to the symmetry of the Jahn–Teller system. Instead, we employ additional HOs. Recall that in Sec. IV C we stated that the number of multivariate Gaussians $\tilde{A}$ comprising $\varrho$ may differ from the number of electronic states $A$ present in the system's Hamiltonian. The Jahn–Teller system has two electronic states, $A = 2$, and the initial choice of sampling distribution, $\varrho_0$, was composed of two multivariate Gaussians $\tilde{A} = 2$. We present two possible alternate choices of $\varrho$, using two additional oscillators ($\varrho_1$), where $\tilde{A} = 4$, and six additional oscillators ($\varrho_2$), where $\tilde{A} = 8$.

We derive $\varrho_1$ from

$$\left(E_i + \hat{h}_o\right)\underline{\mathbb{1}} + \lambda_i \operatorname{diag}\left(+\hat{q}_1, -\hat{q}_1, +\hat{q}_2, -\hat{q}_2\right) \quad (74)$$

and $\varrho_2$ from

$$\left(E_i + \hat{h}_o\right)\underline{\mathbb{1}} + \lambda_i \operatorname{diag}\big(+\hat{q}_1, -\hat{q}_1, +\hat{q}_2, -\hat{q}_2, \\ +\hat{\ell}_1, -\hat{\ell}_1, +\hat{\ell}_2, -\hat{\ell}_2\big), \quad (75)$$

where

$$\hat{\ell}_1 = \frac{\hat{q}_1 + \hat{q}_2}{\sqrt{2}}, \qquad \hat{\ell}_2 = \frac{\hat{q}_1 - \hat{q}_2}{\sqrt{2}}. \quad (76)$$

All $E$, $\omega$, and $\lambda$ values are the same as in Table II. We expect that the increased coverage of the well in $\varrho_1$ and $\varrho_2$ should increase the accuracy of our results. In the bottom two panels of Fig. 4, we see that $\varrho_1$ only has a deviation of $\approx 5\%$, much better than $\varrho_0$'s deviation of $\approx 50\%$. Similarly $\varrho_2$'s deviation appears to be $\leq 1\%$. This shows that changing the placement of a fixed number of distributions $\varrho^a$ is not always sufficient.

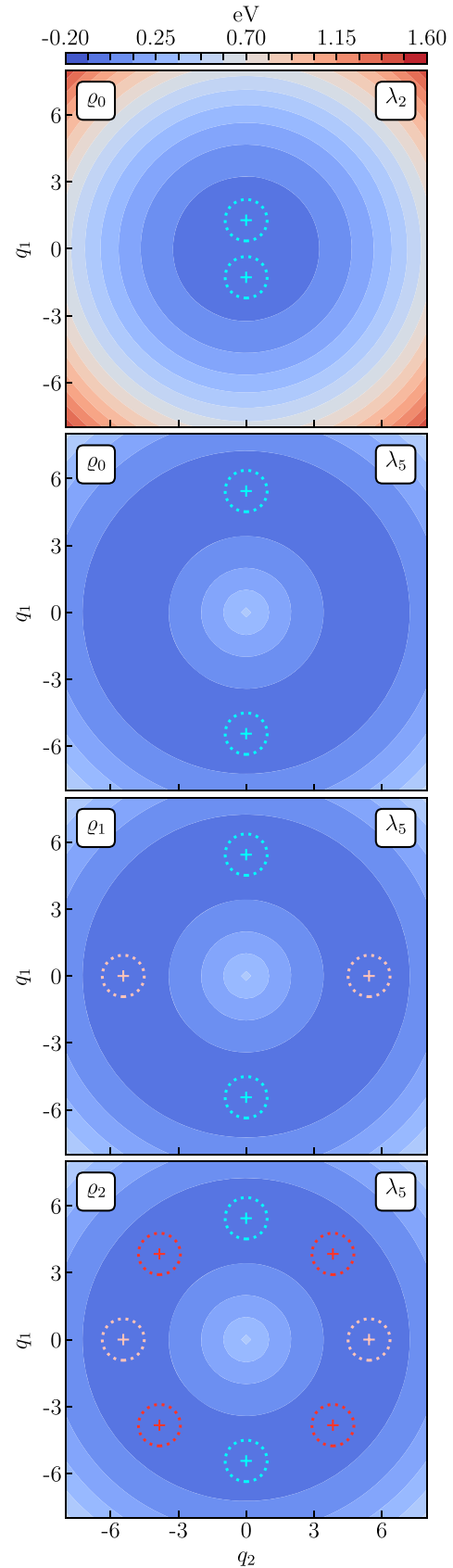

FIG. 3. Four elevation maps of the lower adiabatic PES of the Jahn–Teller system at $\lambda_2 = 0.04$ eV (top panel) and $\lambda_5 = 0.16$ eV (bottom three panels). The colormap for the elevation is presented at the top of the figure. For each panel, the sampling distribution $\varrho_i$ is listed in the upper left-hand corner and an ellipse is centered on the minimum, indicated by the cross (+), of each HO comprising $\varrho_i$ as described in Sec. V.

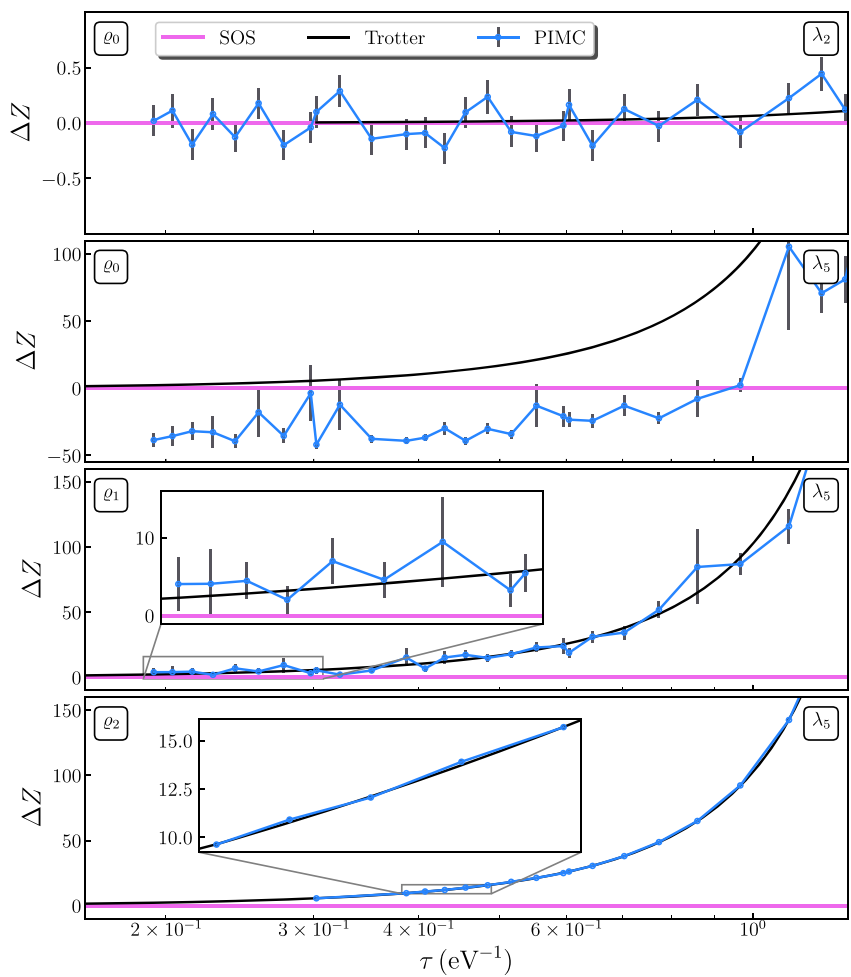

FIG. 4. Four $\tau$ convergence plots of the $\Delta Z$ deviation of our PIMC method (blue) and the exact Trotter (black) from the $Z_{\text{SOS}}$ (magenta). The first panel (from the top) demonstrates that the simple distribution $\varrho_0$ is a reasonable description of the Jahn–Teller system when the curvature of the well is quite small ($\lambda_2$). In the ($\lambda_5$) case, $\varrho_0$ is no longer appropriate as seen in the second panel. Comparing the bottom three panels, it is clear that changing the sampling distribution ($\varrho_0$ to $\varrho_1$ and $\varrho_1$ to $\varrho_2$) has a significantly positive impact on our results.

The number of distributions $\tilde{A}$ must also be considered when choosing $\varrho$ in order to ensure dense coverage of the energy minimum.

## VI. CONCLUDING REMARKS

In this work, we introduced a PIMC method for investigating electronically nonadiabatic systems in thermal equilibrium. Starting from a vibronic Hamiltonian expressed in a product basis of continuous and discrete DoFs, we showed how to arrive at a PI formulation of the canonical partition function $Z$. We derived an expression for $Z$ as the product of a Monte Carlo estimator, which is evaluated stochastically, and a normalization factor, which is evaluated analytically. We also obtained the analytical form of the normalization factor and the distribution $\varrho$ that our algorithm draws samples from. Finally, we presented our algorithm for calculating $Z$.

The importance of choosing an appropriate distribution $\varrho$ was demonstrated. In the testing of model systems, we observed that the accuracy of $\varrho$ as a model of $g$ has a drastic effect on the accuracy of our method. Indeed, we showed that our PIMC method's deviation from exact Trotter calculations is dominated by the choice of $\varrho$.

In the case of our small model systems, obtaining the PESs was computationally feasible and one's intuition was sufficient to guide the selection of the parameters which define $\varrho$. In general, this approach is not feasible and would need to be replaced with a distribution fitting scheme in order to obtain a computationally favourable distribution $\varrho$. As mentioned previously, choosing the GMM is a distribution fitting or parameter estimation problem, for which we can employ common statistical methods such as MLE.[54] The fitting of GMMs is an established area of research in the machine learning community, used in a variety of applications.[63–65] It has been demonstrated that very large mixtures of Gaussians are efficiently learnable in high dimension.[66] We hope to take advantage of these recent advances when applying our PIMC method to more complex systems where the PES cannot be so thoroughly examined. Reasonable convergence in these systems should be attainable with sufficiently well designed GMMs and fitting schemes.

We have built a theoretical and computational framework for obtaining the vibrational-electronic partition function $Z$ for a single molecule described by a vibronic Hamiltonian as illustrated by the applications to our two model systems. This framework can be combined with the usual ideal gas treatment and an approximate rigid rotor model to obtain the molar Helmholtz energy and Gibbs energy $G$. In the future, we aim to develop stable estimators for energetic properties and their fluctuations such as $U$ and $C_v$. Our ultimate goal would be to calculate the $\Delta G$ for chemical reactions where both reactants and products can be described by a vibronic model.

## ACKNOWLEDGMENTS

This research has been supported by the Natural Sciences and Engineering Research Council of Canada (NSERC), the Ontario Ministry of Research and Innovation (MRI), the Canada Foundation for Innovation (CFI), and the Canada Research Chair program.